\documentclass[12 pt, letter]{article}
\usepackage{graphicx}
\usepackage{caption}
\usepackage{amsmath}
\usepackage{amsthm}
\usepackage{enumerate}
\usepackage{subfig}
\usepackage{float}
\usepackage{amsfonts}
\usepackage{amssymb}
\usepackage{anysize}
\usepackage{helvet}
\usepackage{setspace}
\usepackage{multirow}
\usepackage{tabularx}
\usepackage{tabulary}
\usepackage{longtable}
\usepackage{float}
\usepackage{mathtools}
 \def\ds{\displaystyle}
\title{\bf Analytical approximation to the dynamics of a binary stars system with time depending mass variation}
\author{Gustavo V. L\'opez\footnote{gulopez@cencar.udg.mx}~ and Elkin L. L\'opez~\\ \\
 Departamento de F\'{i}sica, Universidad de Guadalajara,\\
 Blvd. Marcelino Garc\'{i}a Barragan y Calzada Ol\'{i}mpica, \\�44200 Guadalajara, Jalisco, Mexico}
\begin{document}
\maketitle

\begin{abstract}
\noindent
We study the classical dynamics of a binary stars when there is an interchange of mass between them. Assuming that one of the star is more massive 
than the other, the dynamics of the lighter one is analyzed as a function of its time depending mass variation. Within our approximations and models for mass
transference, we obtain a general result which establishes that if the lightest star looses mass, its period increases. If the lightest star win mass, its period decreases.  
\end{abstract}
\newpage
\section{ Introduction}
A binary stars system is one of the most common systems in a Galaxy and the Universe [1,2]. This binary system form an effective gravitational potential which bring about five Lagrangian points in the space where gravitational force is zero [3]. one of the most important Lagrangian points is called $L_1$ and is located between the two stars, where the equipotential called Roche-Lobe [4], makes a cross path. If the mass of the star fulfill its lobe, there will be a transference of mass to the other star, through the Lagrange point $L_1$.  This is one of the most common mechanism of transference of mass between two stars , and it is the one we are interested in our study. A usual situation occurs when one star has finished its nuclear cycle fuel and becomes a red giant [5]. Of course, to fully analyze this phenomenon we requires computer simulation of the whole system. However, it would be important if we had a qualitative understanding of the dynamics involved, and this is the propose of this paper. On the first part of our study, the six ordinary second order equations which represents the dynamics of the binary system with time depending mass transference is reduced to a 1-D non autonomous system. This non autonomous system is solved numerically, and the results are analyzed.  
\section{Analytical Approach}
Let $m_1(t)$ and $m_2(t)$ be the masses of the stars in the binary system such that $m_1(t)>m_2(t)$ and $m_1(t)+m_2(t)=M_0$ (constant) for all the time into consideration. This means that we will only consider the transference of mass between the two stars. With respect an inertial reference system (fixed with respect the fixed Galaxies), the non relativistic motion of the system is described by Newton's equation [6]
\begin{subequations}
\begin{equation}
\frac{d m_1(t) {\bf v}_1}{dt}=-G\frac{m_1(t)m_2(t)}{|{\bf x}_2-{\bf x}_1|^2}\widehat{({\bf x}_2-{\bf x}_1)}
\end{equation} 
and
\begin{equation}
\frac{d m_2(t) {\bf v}_2}{d t}=-G\frac{m_1(t)m_2(t)}{|{\bf x}_2-{\bf x}_1|^2}\widehat{({\bf x}_1-{\bf x}_2)}
\end{equation}
\end{subequations}
where ${\bf x}_1$ and ${\bf x}_2$ are the positions, and ${\bf v}_1$ and ${\bf v}_2$ are the velocities of the stars (${\bf v}_i=d{\bf x}_i/dt$). $G$ is the
gravitational constant ($G=6.6738\times 10^{-11}m^3/Kg\cdot s^2$)[7]. The above equations represent six second order ordinary differential equations (6-D problem), and it is not difficult to see that making the usual change of variables to center of mass (${\bf R}=(m_1{\bf x}_1+m_2{\bf x}_2)/M_0$) and relative 
(${\bf x}={\bf x}_2-{\bf x}_1$) coordinates does not help to reduce the number of degree of freedom in our system due to the time depending mass variation. 
However, since we are assuming that $m_1(t)>m_2(t)$, we can select our reference system such that ${\bf x}_1={\bf 0}$ (non inertial reference system seen from the far away fixed-galaxies in the sky), and 
neglect all non inertial forces (Coriolis) which normally appears on this type of reference systems. In addition, our stars are assumed to be points-like mathematical elements. So, making this selection, and defining ${\bf x}={\bf x}_2$ and ${\bf v}={\bf v}_2$, one gets the following Newton's equation of motion for the binary system
\begin{equation}\label{eq1}
\frac{d m_2(t){\bf v}}{dt}=-G\frac{m_1(t)m_2(t)}{r^2}\hat {\bf r},
\end{equation}  
where ${\bf x}=(x,y,z)$, $\hat{\bf r}={\bf x}/|{\bf x}|$, and $|{\bf x}|^2=r^2=x^2+y^2+z^2$. Making the differentiation with respect the time of $m_2$, the equation (\ref{eq1}) can be written as
\begin{equation}\label{eq2}
m_2(t)\frac{d{\bf v}}{dt}=-G\frac{m_1(t)m_2(t)}{r^2}\hat{\bf r}-\dot m_2(t){\bf v}, 
\end{equation}
where $\dot m_2=dm_2/dt$. Let us write now this equation in spherical coordinates,
\begin{equation}
{\bf x}=(r\sin\theta\cos\varphi, r\sin\theta\sin\varphi, r\cos\theta),
\end{equation}
where the position (${\bf x}$), velocity (${\bf v}$)  and acceleration (${\bf a}=d{\bf v}/dt$) are written in term of the unitary vectors $\hat{\bf r}$, $\hat\theta$ and $\hat\varphi$ as
\begin{subequations}
\begin{equation}\label{po}
{\bf x}=r\hat{\bf r},
\end{equation}
\begin{equation}\label{ve}
{\bf v}=\dot r \hat {\bf r}+r\dot\theta \hat\theta+r\dot\varphi\sin\theta \hat\varphi, 
\end{equation}
and
\begin{eqnarray}\label{ac}
{\bf a}&=& (\ddot r-r\dot\theta^2-r\dot\varphi^2\sin^2\theta)\hat{\bf r}+(r\ddot\theta+2\dot r\dot\theta-r\dot\varphi^2\sin\theta\cos\theta)\hat\theta\nonumber\\
& &+(r\ddot\varphi\sin\theta+2\dot r\dot\varphi\sin\theta+2r\dot\varphi\dot\theta\cos\theta)\hat\varphi,
\end{eqnarray}
\end{subequations}
where the unitary vectors are
\begin{subequations}
\begin{eqnarray}
\hat{\bf r}&=& (\sin\theta\cos\varphi,\sin\theta\sin\varphi,\cos\theta),\\
\hat\theta&=&(\cos\theta\cos\varphi,\cos\theta\sin\varphi,-\sin\theta),\\
\hat\varphi&=&(-\sin\varphi,\cos\varphi,0).
\end{eqnarray}
\end{subequations}
Using (\ref{po}), (\ref{ve}), and (\ref{ac}) in (\ref{eq2}), the following equations are obtained
\begin{subequations}
\begin{eqnarray}
&m_2(\ddot r-r\dot\theta^2-r\dot\varphi^2\sin^2\theta)=-G\frac{\ds m_1m_2}{\ds r^2}-\dot m_2\dot r,\\�\nonumber\\
&m_2(r\ddot+2\dot r \dot\theta-r\dot\varphi^2\sin\theta\cos\theta)=-\dot m_2 r\dot\theta,\\�\nonumber\\
&m_2(r\ddot\varphi\sin\theta+2\dot r\dot\varphi\sin\theta+2r\dot\varphi\dot\theta\cos\theta)=-\dot m_2r\dot\varphi\sin\theta.\label{3r}
\end{eqnarray}
\end{subequations}
From (\ref{3r}) we observe that $\dot\varphi=0$ is a possible solution of these equations ($\varphi=\varphi_0=constant$), meaning that that the motion of the system 
can occur in the plane defined by $\varphi=\varphi_0$. Selecting then  this solution, the above equations are reduced to a 2-D problem
\begin{subequations}
\begin{eqnarray}
&m_2(\ddot r-r\dot\theta^2)=-G\frac{\ds m_1m_2}{\ds r^2}-\dot m_2\dot r,\label{1r}\\�\nonumber\\
&m_2(\ddot r+2\dot r \dot\theta)=-\dot m_2 r\dot\theta,\label{2r}
\end{eqnarray}
\end{subequations}
Multiplying (\ref{2r}) by $r$ and rearranging terms, one gets that $d(m_2r^2\dot\theta)/dt=0$ which implies that one obtains the following constant of motion
\begin{equation}\label{theta}
l_{\theta}=m_2r^2\dot\theta.
\end{equation}
finally, using this constant of motion in (\ref{1r}), the study is reduced to the following 1-D problem
\begin{equation}
m_2\ddot r=\frac{\ds l_{\theta}^2}{\ds m_2r^3}-G\frac{\ds m_1m_2}{\ds r^2}-\dot m_2\dot r.
\end{equation}
This equation can be written as the following non autonomous dynamical system
\begin{eqnarray}
\dot r&=& v\\ \nonumber\\
\dot v&=&=\frac{\ds l_{\theta}^2}{m_2^2r^3}-\frac{\ds Gm_1}{\ds r^2} -\frac{\ds \dot m_2}{\ds m_2}v.
\end{eqnarray}
This dynamical system is not integrable, but it can be analyzed numerically. To do this, a model for the mass transference is required, and this will be seen below. 
\section{Models and Results}
We will use the following three models of mass transference 
\begin{eqnarray}
(A)\quad m_2(t)&=&m_{02}-\mu t,\\
(B)\quad m_2(t)&=&m_{02}e^{-\alpha t},\\
(C)\quad m_2(t)&=&m_{02}+m_{01}(1-e^{-\alpha t}), 
\end{eqnarray}
where $\mu$ and $\alpha$ are constant which are relate with the mass lost (win) rate, $m_{01}$ and $m_{02}$ represent the initial mass of the stars ($m_{01}>m_{02}$).  Figures 1, 2 and 3 show the period of the star with mass $m_2$ as a function of the number of turns around the star with mass $m_1$ for the models A, B and C. The initial values taken to make these figures are
\begin{eqnarray}
r(0)&=&456.8965~R_{\odot},\quad v(0)=0~(aphelion),\quad \theta(0)=0,\\
m_1(0)&=&14.14~M_{\odot},\quad m_2(0)=0.85~M_{\odot},\quad l_{\theta}=3.139\times 10^{-35}\frac{M_{\odot}R_{\odot}^2}{year},
\end{eqnarray}
where $R_{\odot}$ and $M_{\odot}$ are the radius and solar mass ($R_{\odot}=6.957\times10^5Km$ and 
$M_{\odot}=1.988\times 10^{30} Kg$ [8]). The values of the parameters $\mu$ and $\alpha$
shown in these figures were chosen big enough to point out the effect of mass variation on the period of the system. These results tell us that if $m_2$ looses mass, its period is larger each turn and, eventually, the star escapes (or breaks) the binary system. On the other hand, if $m_2$ increases  its mass, its period 
becomes smaller each turn. These results are in agreement with what was found on the dynamics of the comets around a star [7].   
\begin{figure}[H]
\includegraphics[scale=0.70,angle=0]{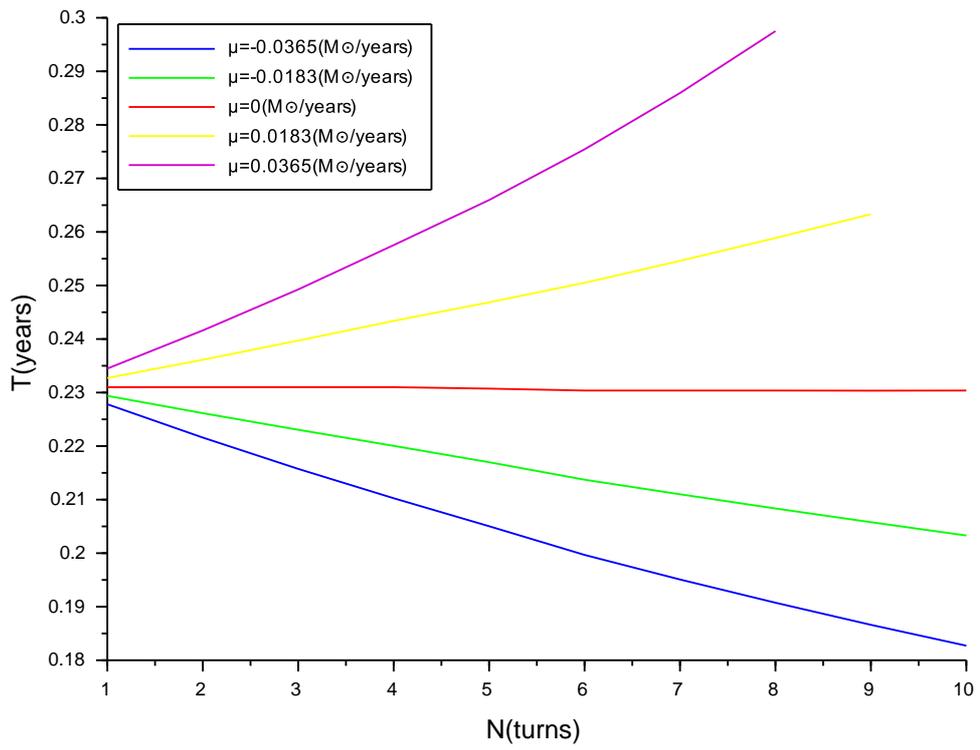}
\centering
    \caption{ Period of $m_2$ with model A.  }
\end{figure}
\begin{figure}[H]
\includegraphics[scale=0.70,angle=0]{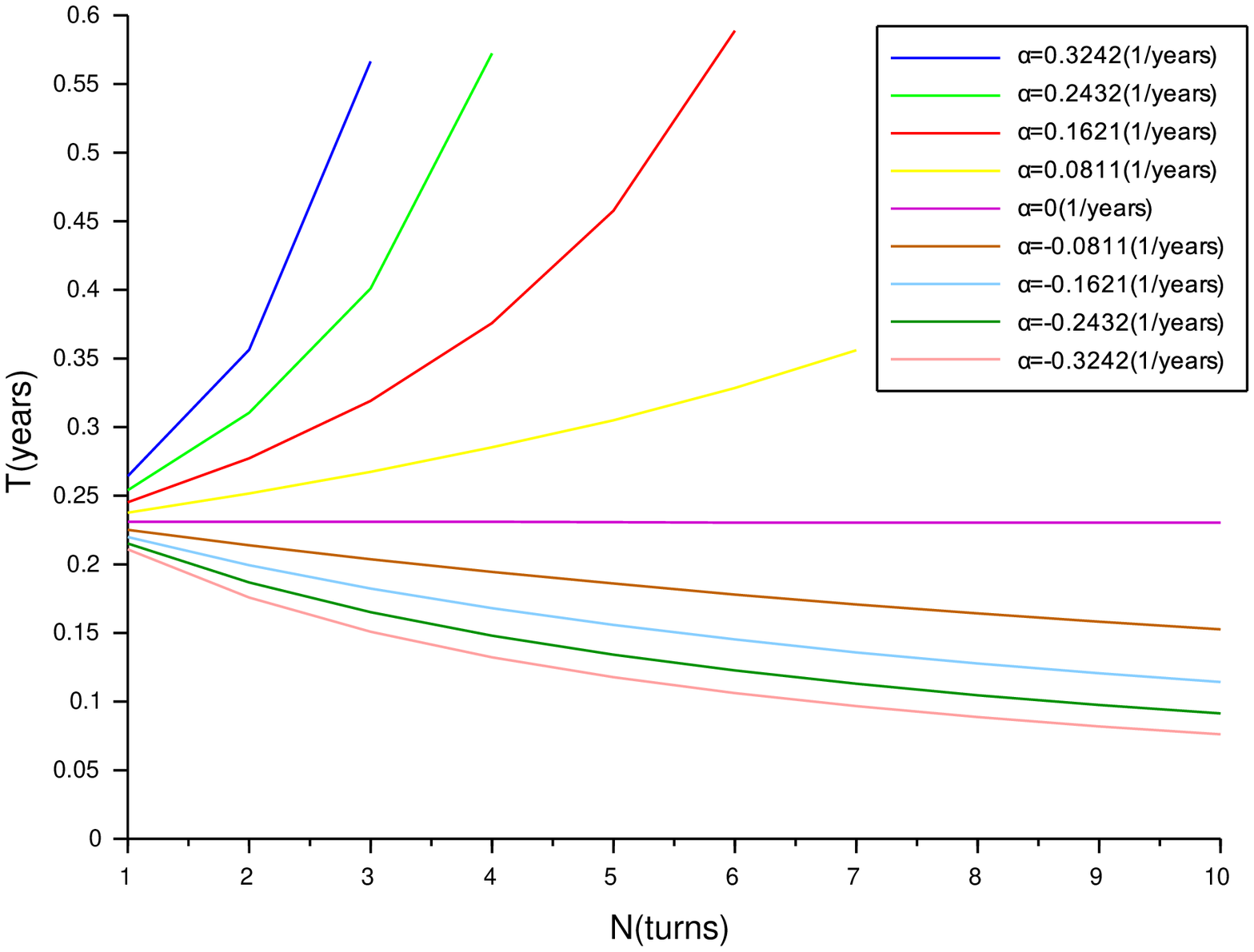}
\centering
    \caption{ Period of $m_2$ with model B.  }
\end{figure}
\begin{figure}[H]
\includegraphics[scale=0.70,angle=0]{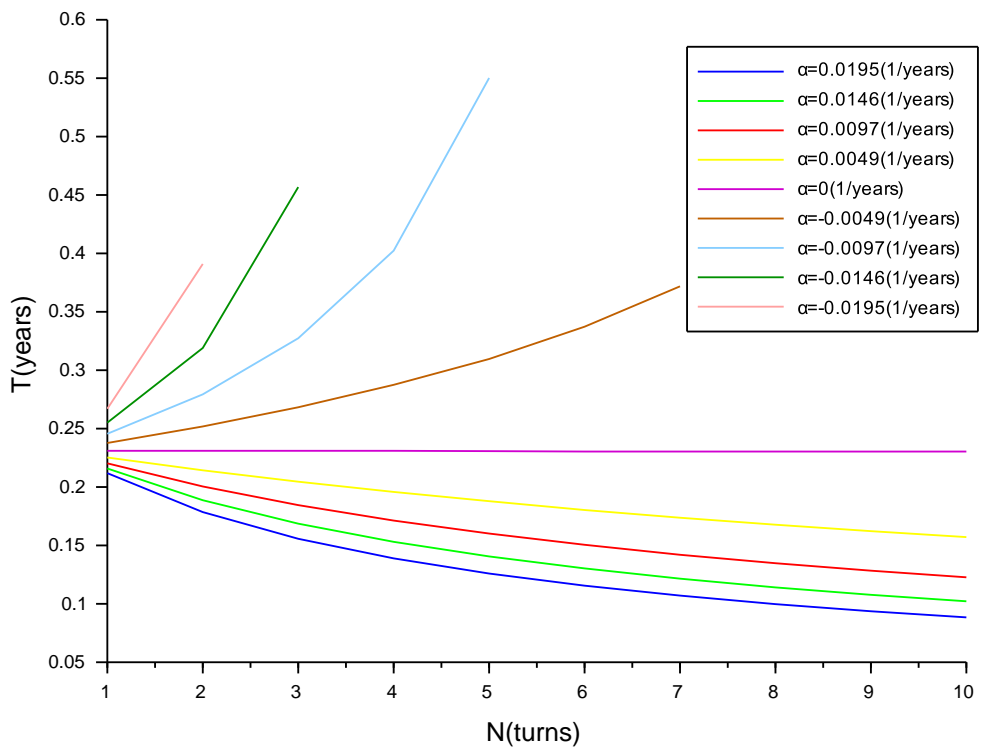}
\centering
    \caption{ Period of $m_2$ with model C.  }
\end{figure} 
\newpage\noindent
\section{Conclusions}
Using several approximations, the dynamics of a class of binary stars with six degrees of freedom with variable mass is reduced to one degree of freedom system
with variable mass. This variation of mass was chosen such that the mass of the whole system remains constant (closed system). Using three models of mass transference between the stars, we found in general that it the less massive star increases its mass from the massive star, its period becomes smaller each turn, and vice versa. The mass rate exchanged  has been taken here too unrealistically large to have better visualization of the expected effect, that is, this effect does not depend on the mass model. The reason is clearly seen from expressions (\ref{eq2}) or the dynamical system (\ref{po},\ref{ve}) since it $\dot m_2$ is positive, one has a damping system, and if $\dot m_2$ is negative, one gets an anti-damping system. 
\newpage\noindent
{\Large\bf References}\\�\\
1. J. Sahade and F.B. Wood, {\it Interacting Binary Stars}, Series in Natural Phylosophy, {\bf 95}, Elsevieir, (2015).\\ \\
2. R.G. Aitken, {\it The Binary Stars}, Dover (1964). P. Gharami, K. Ghosh, and F. Rahaman, {\it A  theoretical Model of Non-conservative Mass Transfer with Non-uniform Mass Accretion Rate in Close Binary Stars}, arXiv:1407.2498v2 [gr-qc], (2014). \\ \\
3. J.L. Lagrange, {\it Essai sur le probl$\grave{e}$m des trois corps}, Gauthier-Villars ($\OE$OVRES DE LAGRANGE), (1873).\\ \\
4. B. Paczynski, Anual Review of Astronomy and Astrophysics, {\bf 9}, (1971), 183. \\
A. Jorissen and A. Frankowski, {\it Detection methods of binary stars with low-and intermediate-mass components}, arXiv:0804.3720v2 [astro-ph], (2008). \\ 
G.J. Savonije, Astron. and AstroPhys., {\bf 62}, (1978), 317.  C.A. Tout and P.P. Eggleton, {\it Tidal enhancement by a binary companion of stellar winds from cool giants}, Osford University Press, (1988).\\ \\
5. C.E. Rolfs and W.S. Rodney, {\it Cauldrons in the Cosmos}, The University of Chicago Press, (1988).\\ \\
6. H. Goldstein, C.Poole, and J. Safko, {\it Classical Mechanics}, Addison Wesley, (2000).\\ \\
7. G.V. L\'opez, J. Mod. Phys. , {\bf 4} (2013), 1638.\\�\\
8.  K.A. Olive et al. (Particle Data Group), Chinese Physics C{\bf 38}, 090001 (2014).

\end{document}